\documentclass[aps,notitlepage,nofootinbib]{revtex4-1}

\usepackage{amstext,amsmath,amssymb,amsfonts,bbm}
\usepackage[latin1]{inputenc}
\usepackage{epsfig}
\usepackage{hyperref}
\usepackage{amsthm}
\usepackage{subfigure}
\usepackage{color}
\usepackage{multirow}

\newcommand{\bea}{\begin{eqnarray}}	
\newcommand{\eea}{\end{eqnarray}}
\newcommand{\be}{\begin{equation}}	
\newcommand{\ee}{\end{equation}}
\newcommand{\cG}{{\cal G}}

\newcommand{\cT}{{\cal T}}

\newcommand{\cF}{{\cal F}}
\newcommand{\cB}{{\cal B}}
\newcommand{\cL}{{\cal L}}
\newcommand{\cN}{{\cal N}}
\newcommand{\cJ}{{\cal J}}
\newcommand{\cH}{{\cal H}}
\newcommand{\cR}{{\cal R}}

\newtheorem{definition}{Definition}
\newtheorem{theorem}{Theorem}

\begin{document}

\title{\large \bf The $1/N$ expansion of colored tensor models}

\author{Razvan Gurau}\email{rgurau@perimeterinstitute.ca}
\affiliation{Perimeter Institute for Theoretical Physics \\ 31 Caroline
St, Waterloo, Ontario N2L 2Y5, Canada}

\begin{abstract}
\noindent In this paper we perform the $1/N$ expansion of the colored three dimensional 
Boulatov tensor model. As in matrix models, we obtain a systematic topological expansion, with 
increasingly complicated topologies suppressed by higher and higher powers of $N$. 
We compute the first orders of the expansion and prove that only graphs 
corresponding to three spheres $S^3$ contribute to the leading order 
in the large $N$ limit.
\end{abstract}

\maketitle

\section{Introduction}

Random tensor models and Group Field Theories (GFT) \cite{laurentgft,quantugeom2} 
generalize random matrix models \cite{mm,mmgravity,ambj3dqg,sasa1} 
to higher dimensions. The Feynman graphs of GFT are built from vertices dual to
$n$ simplices, and propagators encoding the gluing of $n$ simplices along 
boundary $(n-1)$ simplices. Parallel to ribbon graphs of matrix models 
(dual to discretized surfaces), GFT graphs are dual to discretized 
$n$ dimensional topological spaces. For the simplest GFT models \cite{GFT} the Feynman 
amplitude of a graph, reproduces the partition function of discretized BF theory  
\cite{FreidelLouapre,gftnoncom}\footnote{More involved GFT models
\cite{newmo2, newmo3, newmo4, newmo5} have been proposed in an attempt 
to implement the Plebanski constraints and reproduce the gravity partition function.}.

In contrast with random matrix models, the usual GFT models suffer from two major problems. 
First the Feynman graphs of GFTs are dual
not only to manifolds and pseudo manifolds but also to more singular spaces
\cite{sing}. Second, and even more problematic, no equivalent of the $1/N$ expansion 
or of the notion of planarity \cite{Brezin:1977sv} crucial in matrix models 
had yet been found in GFTs. This is one of the most important challenges GFTs and
tensor models face today \cite{Alexandrov:2010un}. 
The recently introduced ``colored GFTs'' \cite{color,PolyColor} (CGFT)
solve the first problem \cite{sing} and generate only pseudo manifolds. In this paper
we prove that they also solve the second problem, namely CGFTs admit a $1/N$ 
topological expansion. We present in this paper the systematic expansion at all orders 
of CGFT and explicitly compute the first terms. We prove that at leading order 
in $1/N$ only graphs dual to the three sphere $S^3$ contribute. 
To establish this result we will rely on one hand on results and methods 
introduced in \cite{FreiGurOriti,sefu1,sefu2,sefu3, param} concerning 
amplitudes of CGFT graphs and on the other on results in combinatorial topology 
and manifold crystallization theory \cite{Lins, FG}. Almost none of
the concepts and techniques we use can be applied to 
non colored GFT models.

This paper is organized as follows. In Section \ref{sec:model} we recall
the colored three dimensional Boulatov tensor model. Sections
\ref{sec:ribbon} and \ref{sec:dipoles} introduce the techniques 
required to perform in section \ref{sec:multigems} the $1/N$ expansion 
of the model.

\section{The Colored Boulatov Model}\label{sec:model}

Let $G$ be some compact multiplicative Lie group, and denote $h$ its elements,
$e$ its unit, and $\int dh$ the integral with respect to the Haar measure.
Let $\bar \psi^i,\psi^i$,  $i=0,1,2,3$ be four couples of complex 
scalar (or Grassmann) fields over three copies of $G$, $\psi^i:G\times G\times G 
\rightarrow \mathbb{C}$. We denote $\delta^N(h)$ the 
delta function over $G$ with some cutoff such that 
$\delta^N(e)$ is finite, but diverges (polynomially) when $N$ goes to infinity. For 
$G=SU(2)$ (denoting $\chi^j(h)$ the character of $h$ in the representation $j$) 
respectively $G=U(1)$ we can chose 
\bea
  \delta^N(h)\Big{|}_{G=SU(2)} = \sum_{j=0}^{N} (2j+1) \chi^{j}(h) \qquad
\delta^N(\varphi)\Big{|}_{G=U(1)} = \sum_{p=-N}^{N} e^{\imath p \varphi} \; .
\eea

The partition function of the colored Boulatov model \cite{color} over $G$ is the path integral
\bea\label{eq:part}
  Z(\lambda,\bar\lambda)= e^{-F(\lambda,\bar\lambda)} = 
\int \prod_{i=0}^4 d\mu_P(\bar \psi^i,\psi^i) \; e^{-S^{int}(\bar \psi^i,\psi^i)} \; ,
\eea
with normalized Gaussian measure of covariance $P$
\bea
&& P_{h_{0}h_{1}h_{2} ; h_{0}'h_{1}'h_{2}'} = 
\int d\mu_P(\bar \psi^i,\psi^i) \; \bar\psi^i_{h_{0}h_{1}h_{2}}
  \psi^i_{h_{0}'h_{1}'h_{2}'} 
 = \int dh \;  \delta^{N}\bigl( h_{0} h (h_{0}')^{-1} \bigr)
\delta^{N}\bigl( h_{1} h (h_{1}')^{-1} \bigr)
\delta^{N}\bigl( h_{2} h (h_{2}')^{-1} \bigr) \; ,
\eea
and interaction (denoting $\psi(h,p,q)=\psi_{hpq}$)  
\bea\label{eq:interaction}
S^{int}&=& \frac{\lambda}{\sqrt{\delta^N(e)}} 
\int_{G^6} \psi^0_{h_{03}h_{02}h_{01}} 
 \psi^1_{h_{10}h_{13}h_{12}} \psi^2_{h_{21}h_{20}h_{23}} 
 \psi^3_{h_{32}h_{31}h_{30}} \crcr
&&+\frac{\bar\lambda}{\sqrt{\delta^N(e)}} 
\int_{G^6} \bar \psi^0_{h^{03}h^{02}h^{01}} 
 \bar \psi^1_{h^{10}h^{13}h^{12}} \bar \psi^2_{h^{21}h^{20}h^{23}} 
 \bar \psi^3_{h^{32}h^{31}h^{30}} \; ,
\eea
where $h_{ij}=h_{ji}$. We call ``black'' the vertex involving the $\psi$'s and
``white'' the one involving the $\bar \psi$'s. 

\begin{figure}[htb]
\begin{center}
\includegraphics[width=4cm]{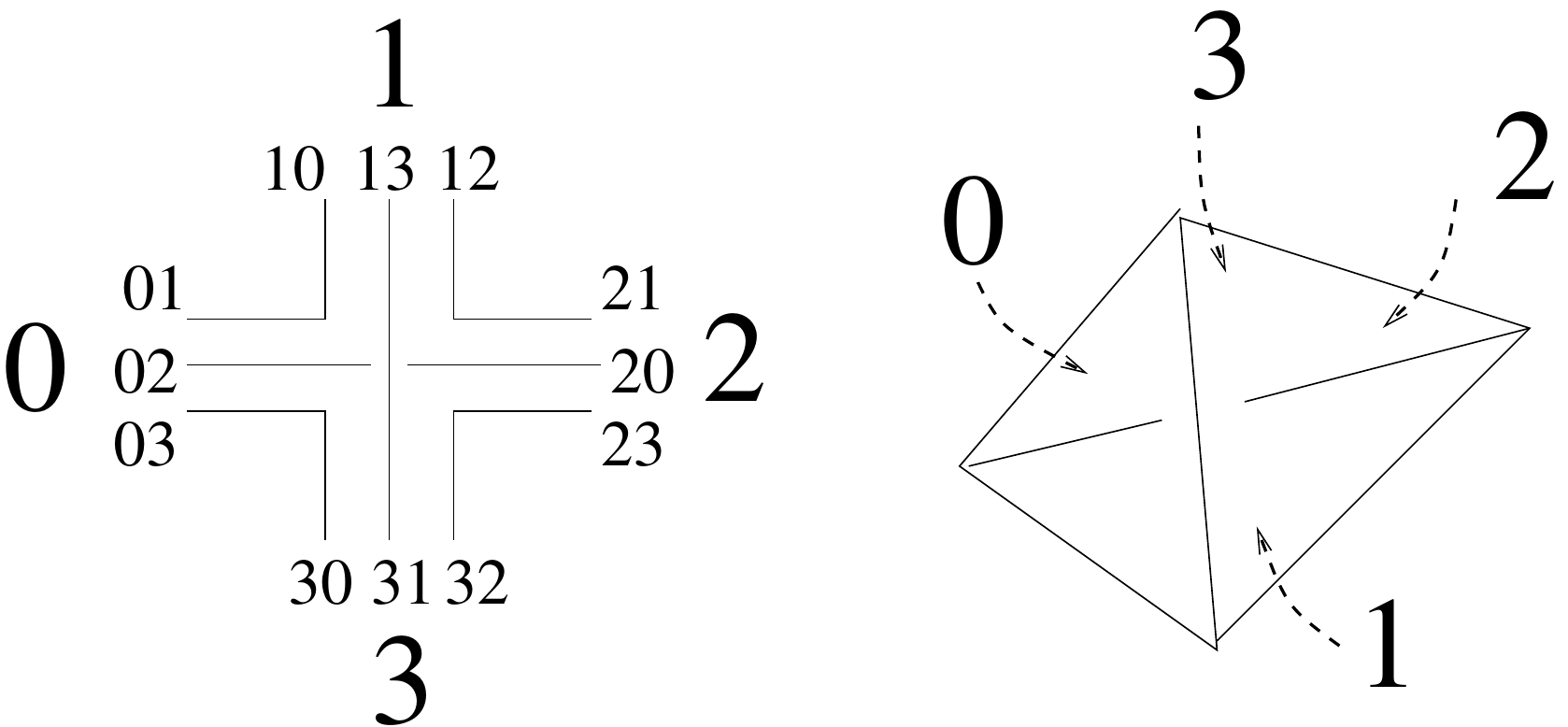}\hspace{0.5cm} 
\includegraphics[width=4cm]{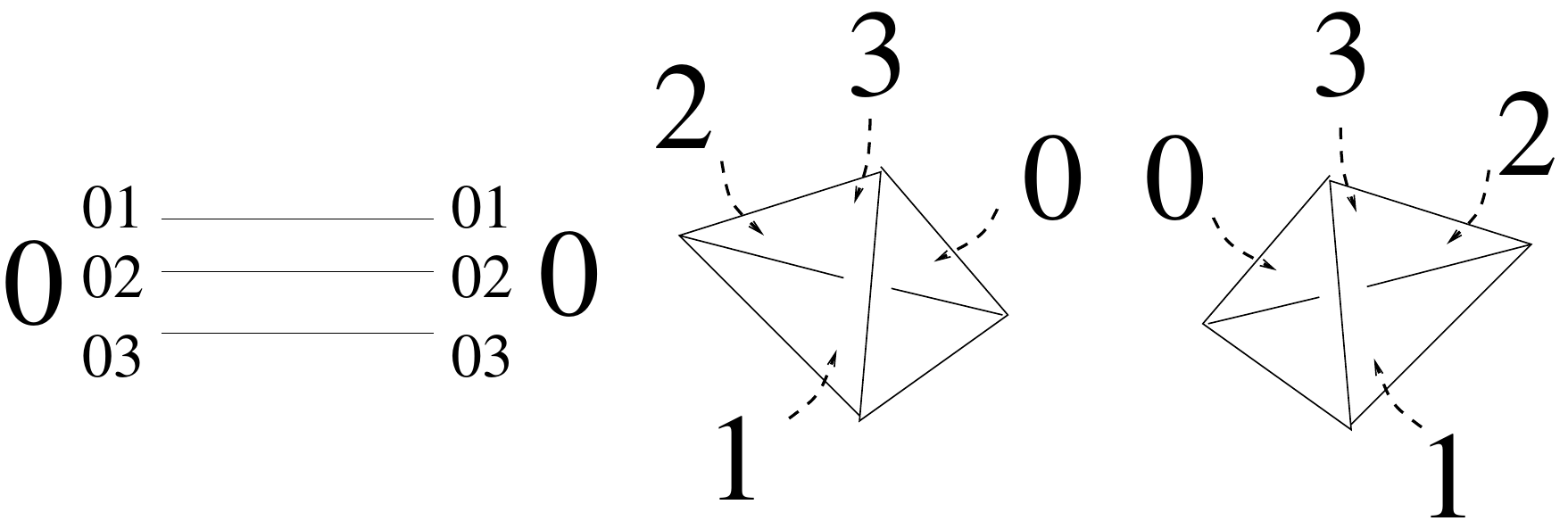}
\caption{Colored GFT vertex and line.}
\label{fig:vertex}
\end{center}
\end{figure}

The half lines of the CGFT vertex (represented in figure \ref{fig:vertex}) have a color $i$. 
The group elements $h_{ij}$ in eq. (\ref{eq:interaction}) are associated to the  ``strands'' (represented 
as solid lines) common to the half lines $i$ and $j$. The vertex is dual to a tetrahedron and 
its half lines represent the triangles bounding the tetrahedron. The strand $ij$, common to the half lines 
$i$ and $j$, represents the edge of the tetrahedron common to the triangles $i$ and $j$. 
The CGFT lines (figure \ref{fig:vertex}) always connect two
vertices of opposite orientation (i.e. a black and a white vertex). 
They have three {\it parallel} strands 
associated to the three arguments of the fields. A line represents the gluing of 
two tetrahedra (of opposite orientations) along triangles of the same color.

\begin{figure}[htb]
\begin{center}
 \includegraphics[width=3.5cm]{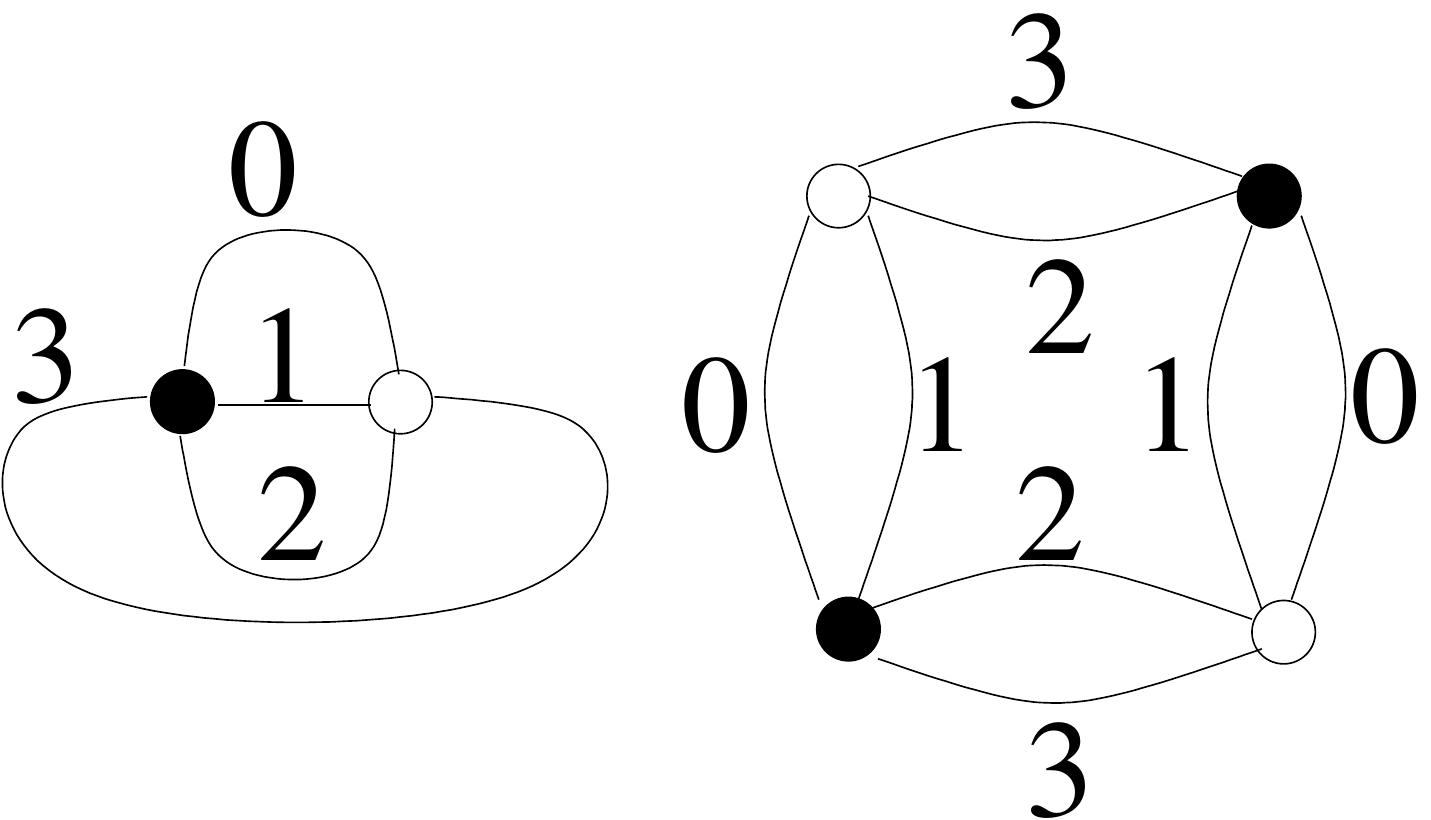} \hspace{1cm}
 \includegraphics[width=3.5cm]{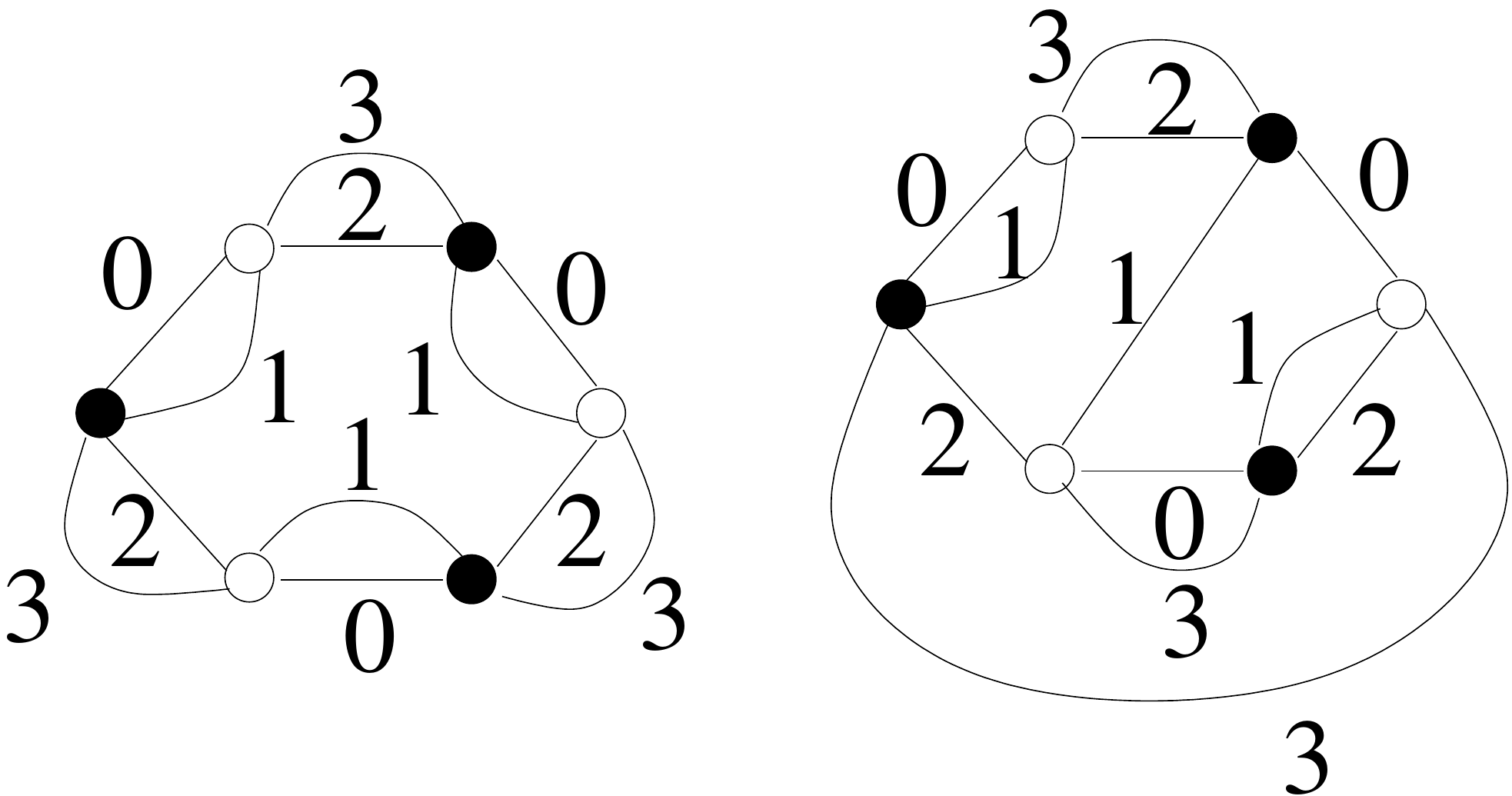} \hspace{1cm}
 \includegraphics[width=3.5cm]{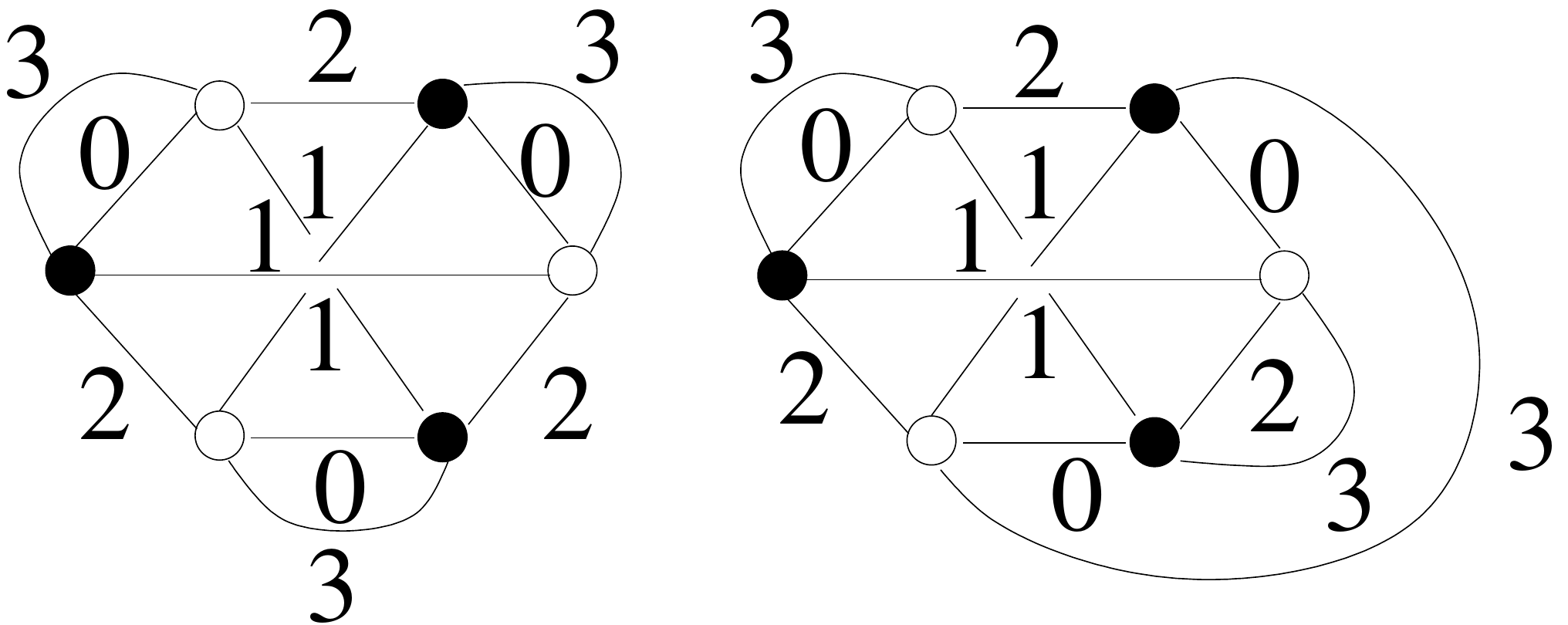} 
\caption{Examples of Colored GFT graphs.}
\label{fig:graphs}
\end{center}
\end{figure}

The strand structure of the vertex and propagator is fixed. One
can represent a CGFT graph either as a stranded graph (using the vertex and propagator
in figure \ref{fig:vertex}) or as a ``colored graph'' with (colored) solid lines, 
and two classes of oriented vertices. Some examples of CGFT graphs are given 
in figure \ref{fig:graphs}. We denote them from left to right $\cG_1$, $\cG_2$, $\cG_{3;a}$, $\cG_{3;b}$,
$\cG_{3;c}$ and $\cG_{3;d}$.

The lines of a vacuum CGFT graph $\cG$ are oriented (say from the black 
to the white vertex). The closed strands of $\cG$ form ``faces'' and are labeled by 
couples of colors. A vacuum CGFT graph must have the same number
of black and white vertices. In this paper we will only deal with connected graphs.
We denote $\cN_{\cG}$, $\cL_{\cG}$, $\cF_{\cG}$ 
the sets of vertices, lines and faces of $\cG$. Also, we denote $\cL_{\cG}^i$ the set of lines of 
color $i$ and $\cF^{ij}_{\cG}$ the set of faces of colors $ij$ of $\cG$. The Feynman amplitude of 
$\cG$ is  
\bea\label{eq:ampli}
 A^{\cG}  = \frac{(\lambda\bar\lambda)^{\frac{\cN_{\cG}}{2}}}{[\delta^N(e)]^{\frac{|\cN_{\cG}|}{2}}} 
\int \prod_{\ell\in \cL_{\cG}} dh_{\ell} 
\prod_{f\in \cF_{\cG}} \delta^N_{f}(\prod_{\ell\in f }^{\rightarrow} h_{\ell}^{\sigma^{ \ell | f}} )\; ,
\eea
where the notation $\ell\in f$ (which we sometimes omit) signifies that the 
line $\ell$ belongs to the face $f$ and $\sigma^{\ell|f}=1$ (resp. $-1$)
if the orientations of $\ell$ and $f$ coincide (resp. are opposite). The $\delta^N$ functions are invariant
under cyclic permutations and conjugation of their arguments hence the amplitude of a graph does not depend on the
orientation of the faces or on their starting point. 

The first ingredient in our $1/N$ expansion is the scaling of the coupling in eq. (\ref{eq:interaction}).
In \cite{sefu2} it is proved that $A^{\cG}$ obeys
\bea
 A^{\cG} \le \frac{(\lambda\bar\lambda)^{ \frac{|\cal N_{\cG}|}{2} }}{[\delta^N(e)]^{ \frac{|\cal N_{\cG}|}{2} }} 
[\delta^{N}(e)]^{\frac{|\cal N_{\cG}|}{2}+2}
=(\lambda\bar\lambda)^{\frac{|\cal N_{\cG}|}{2}} [\delta^{N}(e)]^2 \; ,
\eea
and that the bound is optimal (that is there exist graphs at any order saturating it). 
In order to obtain a sensible large $N$ limit, the scaling of the couplings $\lambda$ and  $\bar \lambda$ 
must be chosen such that the maximally divergent graphs have uniform degree of divergence at all orders. 

\section{Ribbon Graphs}\label{sec:ribbon}

To any CGFT graph one associates two classes of ribbon graphs: its bubbles \cite{color} and its jackets
\cite{sefu3}. We denote in the sequel $\widehat{i} = \{0,1,2,3\}\setminus \{i\}$,
 $\widehat{ij} = \{0,1,2,3\} \setminus \{i,j\}$ and $\widehat{ijk} = \{0,1,2,3\} \setminus \{i,j,k\}$.

\bigskip

\begin{figure}[htb]
\begin{center}
 \includegraphics[width=8cm]{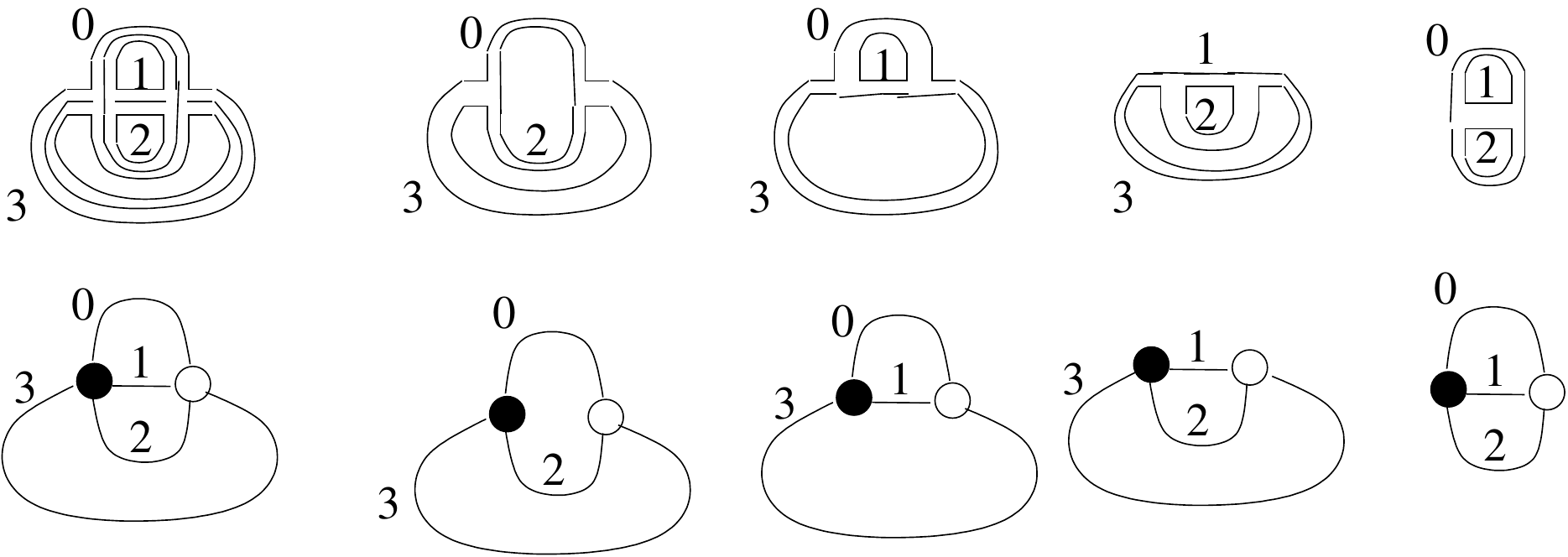}
\caption{The bubbles of $\cG_1$.}
\label{fig:bub}
\end{center}
\end{figure}

\noindent{\bf Bubbles.} 
The bubbles \cite{color} of a CGFT graph are the maximally connected subgraphs with three 
colors. They are dual to the vertices of the gluing of tetrahedra\footnote{
Recently an alternative definition for bubbles has been proposed in \cite{Bonzom:2010ar}.
Although interesting in itself, this definition is somewhat
idiosyncratic, and it seems preferable to use the more standard 
notion of bubbles dual to vertices of the gluing of
tetrahedra.}. The bubbles admit two representations, either as colored 
graphs or as ribbon graphs \cite{color,PolyColor}. 
The ribbon graph of a bubble with colors $i,j,k$ is obtained by deleting all the strands containing 
the color $\widehat{ijk}$. The bubbles of the graph $\cG_1$ (figure \ref{fig:graphs})
are represented in figure \ref{fig:bub}. 
We denote $\cB_{\cG}$ the set of all the bubbles of $\cG$ and $\cB^{ijk}_{\cG}$ the set of 
bubbles of colors $ijk$.

For a bubble $b\in \cB_{\cG}$, we denote $n_b$, $l_b$ and $f_b$ the sets of its 
vertices, lines and faces. 
The graph $\cG$ has four valent vertices ($2|\cN_{\cG}| = |\cL_{\cG}|$), 
while its bubbles have three valent vertices ($3|n_b| = 2|l_b|$). We have
\bea\label{eq:eule}
&& 4|\cN_{\cG}| = \sum_{b\in\cB_{\cG}} |n_b|\; , \quad 3|\cL_{\cG}| =  \sum_{b\in \cB_{\cG}} |l_b| 
\; , \quad 2|\cF_{\cG}| = \sum_{b\in \cB_{\cG} } |f_b|\; , \crcr
&& |\cN_{\cG}|-|\cL_{\cG}|+|\cF_{\cG}| -|\cB_{\cG}| = -\sum_{b\in \cB_{\cG} } g_b \; ,
\eea
with $g_b$ the genus of the bubble $b$. A graph $\cG$ is dual to an orientable pseudo 
manifold. If all its bubbles are planar then it is dual to an orientable manifold \cite{sing}.

\bigskip

\begin{figure}[htb]
\begin{center}
 \includegraphics[width=6cm]{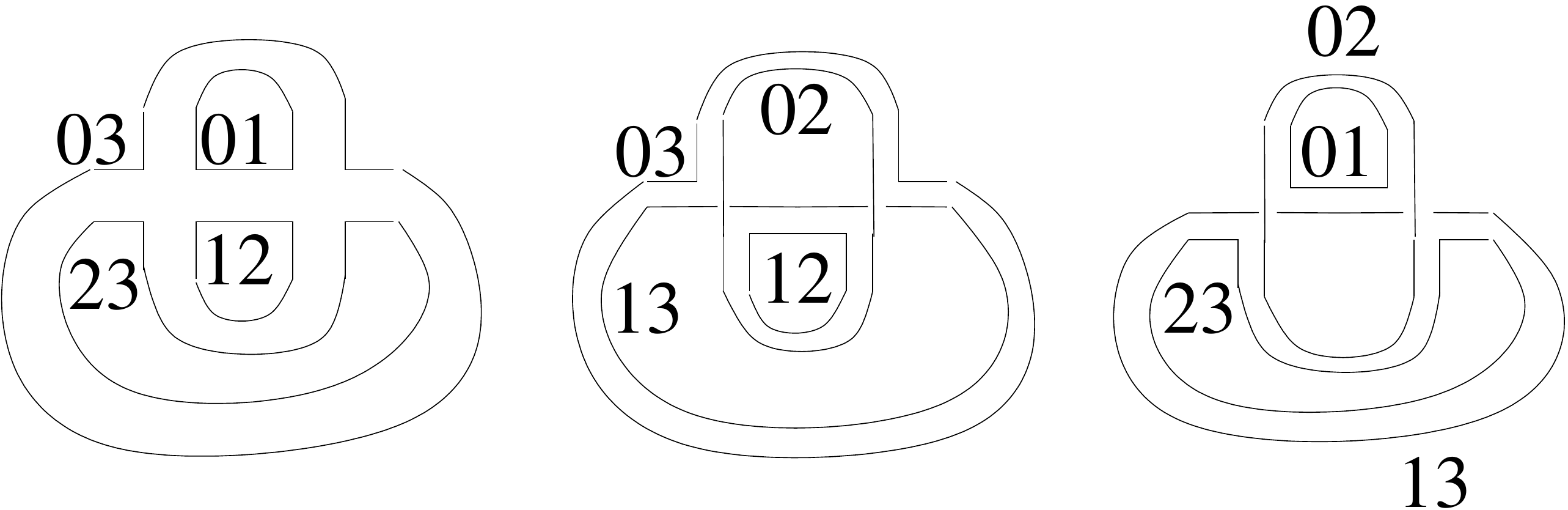}
\caption{The jackets of $\cG_1$.}
\label{fig:jacket}
\end{center}
\end{figure}

\noindent{\bf Jackets.} 
A second class of ribbon graph associated to $\cG$ are its jackets \cite{sefu3}. 
A jacket of $\cG$ is the ribbon graph obtained from $\cG$ by deleting all the faces 
with colors $ij$ and $\widehat{ij}$. 
A CGFT graph has three jackets. The three jackets of $\cG_1$ are represented in figure \ref{fig:jacket}, where
the labels are associated to the faces.

The jackets of $\cG$ have four valent ribbon vertices. The reader might be worried that, while the vertices
of the jacket with faces $02,13$ deleted (the one originally identified in \cite{sefu3}) are simple ribbon vertices, 
the ones of the other two jackets (with the faces $03,12$ and $01,23$ deleted) appear twisted in figure \ref{fig:jacket}. 
This is just an illusion: permuting the half lines $0$ and $1$ and respectively $1$ and $2$ on every jacket vertex 
eliminates all the twists. The sets of vertices, lines and faces of a jacket are $\cN_{\cG}$, $\cL_{\cG}$
and $\cF_{\cG} \setminus \cF^{ij}_{\cG} \setminus \cF^{\widehat{ij} }_{\cG}$.

\bigskip

\noindent{\bf Face routing.} In non identically distributed matrix models \cite{GW,GW1,GW2} the amplitude of 
a Feynman graph is computed via a ``routing'' algorithm, a digested version of which we 
present below.

To every ribbon graph $\cH$ (with sets of vertices, lines and faces denoted $\cN$, $\cL$ and $\cF$) 
one associates a dual graph $\tilde \cH$. The construction is standard (see for instance \cite{param,GW2} 
and references therein). The vertices of $\cH$, correspond to the faces of $\tilde \cH$, 
its lines to the lines of $\tilde \cH$ and its faces to the vertices of $\tilde \cH$.
The lines of $\cH$ admit (many) partitions in three disjoint sets: a tree $\cT$ in $\cH$, 
($|\cT|=|\cN|-1$), a tree $\tilde \cT$ in the its dual $\tilde \cH $, ($|\tilde \cT|=|\cF|-1$), 
and a set $\cL \setminus \cT \setminus \tilde \cT$, ($| \cL \setminus \cT \setminus \tilde \cT|=2g_{\cH}$ ) 
of ``genus'' lines (\cite{param}). 

\begin{figure}[htb]
\begin{center}
 \includegraphics[width=4cm]{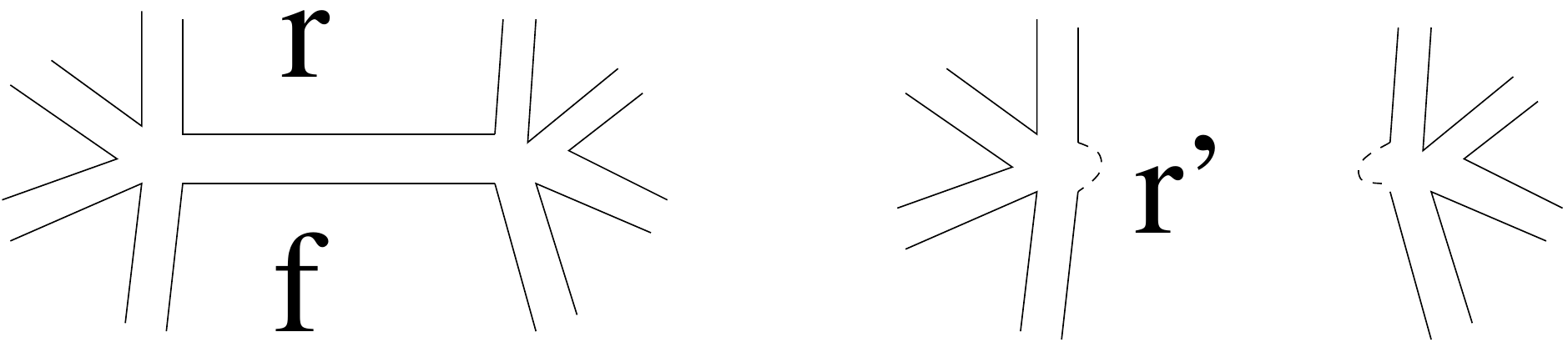}
\caption{Deletion of a ribbon line.}
\label{fig:del}
\end{center}
\end{figure}

We orient the faces of $\cH$ such that the two strands of every line have opposite orientations.
We set a face of $\cH$ as ``root'' (denoted $r$). Consider a faces $f$ sharing some line 
$l(f,\tilde \cT) \in \tilde \cT$ with the root (that is the two strands of $l(f,\tilde \cT)$ belong one 
to $r$ and the other to $f$). The group element $h_{l(f,\tilde \cT)}$ appears exactly once in the argument of
$\delta^N_f$ and $\delta^N_r$ 
\bea\label{eq:rout1}
\delta^N_{r}(\prod_{\ell}^{\rightarrow} h_{\ell}^{\sigma^{\ell | r}}) \;
\delta^N_{f}(\prod_{\ell}^{\rightarrow} h_{\ell}^{\sigma^{\ell | f}})
=\delta^N_{r}\Big{(} (\prod_{\ell \neq l(f,\tilde \cT) }^{\rightarrow} h_{\ell}^{\sigma^{\ell | r}} ) 
h_{l(f,\tilde \cT) }^{\sigma^{ l(f,\tilde \cT) | r}} \Big{)} \;
\delta^N_{f} \Big{(} h_{ l (f,\tilde \cT) }^{\sigma^{l(f,\tilde \cT) | f}} 
(\prod_{\ell \neq l(f,\tilde \cT) }^{\rightarrow} h_{\ell}^{\sigma^{\ell | f}}) \Big{)} \; ,
\eea
where we set $l(f,\tilde \cT) $ as the last line of $r$ and as the first line of $f$. 
By our choice of orientations $\sigma^{l(f,\tilde \cT) | r }\sigma^{l(f,\tilde \cT) | f}=-1 $ and
eq. (\ref{eq:rout1}) becomes
\bea
 \delta^N_{r}\Big{(} (\prod_{\ell \neq l(f,\tilde \cT) }^{\rightarrow} h_{\ell}^{\sigma^{\ell | r}} )
(\prod_{\ell \neq  l(f,\tilde \cT) }^{\rightarrow} h_{\ell}^{\sigma^{\ell | f}} ) 
\Big{)} \;
\delta^N_{f} \Big{(} h_{l(f,\tilde \cT)}^{\sigma^{l(f,\tilde \cT) | f}} 
(\prod_{\ell \neq l(f,\tilde \cT) }^{\rightarrow} h_{\ell}^{\sigma^{\ell | f}}) \Big{)} \; .
\eea
This trivial multiplication has two consequences. First the face $f$ is canonically associated 
to the line $l(f,\tilde \cT)$. Second, the face $r$ becomes a root face in the graph 
$\cH-l(f,\tilde \cT)$, obtained from $\cH$ by deleting $l(f,\tilde \cT)$ and connecting 
$r$ and $f$ into a new face $r'=r \cup f$ (see figure \ref{fig:del}).
Iterating for all faces except the root we get
\bea \label{eq:routing}
 \prod_{f\in \cH} \delta^N_{f}( \prod_{\ell}^{\rightarrow} h_{\ell}^{\sigma^{\ell | f}}) = 
 \delta^N_{r} (\prod_{\ell \notin \tilde \cT }^{\rightarrow} h_{\ell}^{\sigma^{\ell | \cup_{f\in \cH} f} } )
 \prod_{f\in \cH, f\neq r} 
\delta^N_{f} \Big{(} h_{l(f,\tilde \cT)}^{\sigma^{l(f,\tilde \cT) | f}} 
(\prod_{\ell \neq l(f,\tilde \cT) }^{\rightarrow} h_{\ell}^{\sigma^{\ell | f}}) \Big{)} \; .
\eea

If $\cH$ is planar, $\cup_{f\in \cH} f$ is the exterior face of the tree $\cT$ in $\cH$. The group 
elements corresponding to lines of $\cT$ touching leafs (vertices of coordination 
one in $\cT$) appear consecutively $h_lh_l^{-1}$, and drop from the root face. Iterating for all line in $\cT$
we get 
\bea\label{eq:routingf}
 \delta^N_{r} (\prod_{\ell \notin \tilde \cT }^{\rightarrow} h_{\ell}^{\sigma^{\ell | \cup_{f\in \cH} f} } ) 
\Big{\vert}_{\cH \text{ planar}}
   = \delta^N(e) \; ,
\eea
for {\it any} base group $G$. Remark that only the argument of the root $\delta^N_r$ changes
under routing.

\bigskip

\noindent{\bf An Example: $2D$ GFT.} The $2$ dimensional GFT (with $G=SU(2)$) is a non identically
distributed matrix model. The couplings do not need to be rescaled in this case.
The free energy admits a familiar ``genus expansion''.

By face routing one can integrate all group 
elements $h_l$ with $l\in \tilde \cT$ and by a tree change of variables
\cite{FreiGurOriti} one eliminates all group elements $h_l$ with $l \in \cT$. One is left with an integral over the genus
lines corresponding to a ``super rosette graph'' \cite{param} with only one vertex and one face.
The super rosette is obtained from $\cG$ by deleting the lines in $\tilde \cT$ and contracting the lines
in $\cT$. The particular super rosette to which a graph is reduced depends on the routing trees $\cT$ and 
$\tilde \cT$, but all super rosettes associated to a graph have the same genus $g$. 
One can define $[R_g]$ as the equivalence class of all super rosettes of genus $g$.
For a super rosette, each genus line appears twice in the argument of the last $\delta^N$ function. 
We expand in characters and integrate the genus lines (by the ``third Filk move''
in the dual super rosette \cite{param}). Each genus line brings 
a factor $(2j+1)^{-1}$, hence the amplitude of a super rosette is 
$A^{R_g}= \sum (2j+1)^{1-2g}\approx N^{2-2g}$, for all super rosettes of genus $g$\footnote{To correctly identify
the scaling with $N$ one must use sliced $\delta^N$ functions, $\delta^N(h)=\sum_{N/2}^N (2j+1) \chi^j(h)$. }. 
The amplitude of $\cG$ equals the one of the super rosette class 
to which it belongs. The genus expansion of the free energy writes
\bea
 F(\lambda,\bar\lambda) = \sum C^{[R_g]}(\lambda,\bar \lambda) A^{[R_g]}  
= \sum C^{[R_g]}(\lambda,\bar \lambda) \; N^{2-2g}\; ,
\eea
with $C^{[R_g]}(\lambda,\bar \lambda)$ a combinatorial factor counting the graphs which reduce to 
the super rosette class $[R_g]$ i.e. 
all graphs of genus $g$. Of course in $2$ dimensions, as the super rosette amplitudes can be computed explicitly
one completely forgets about them, indexes the expansion of the free energy by the genus $g$ and concludes 
that higher and higher genus graphs are suppressed by larger and larger powers of the cut off. 

\section{Dipoles}\label{sec:dipoles}

The second ingredient we need to establish our results are the Dipole moves \cite{Lins, FG}
encoding homeomorphisms of pseudo manifolds (we will make a precise statement later).
We will identify the various bubbles, faces and lines below by their
colors (in superscript) and their vertices (in subscript).

\bigskip

\begin{figure}[htb]
\begin{center}
 \includegraphics[width=4cm]{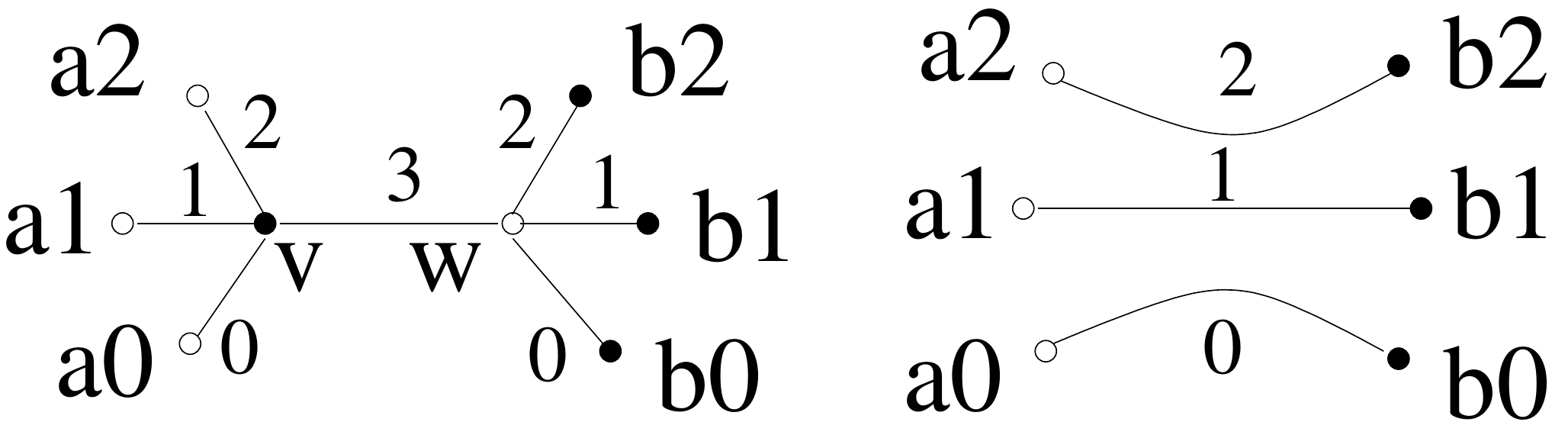} \hspace{.5cm}
 \includegraphics[width=3cm]{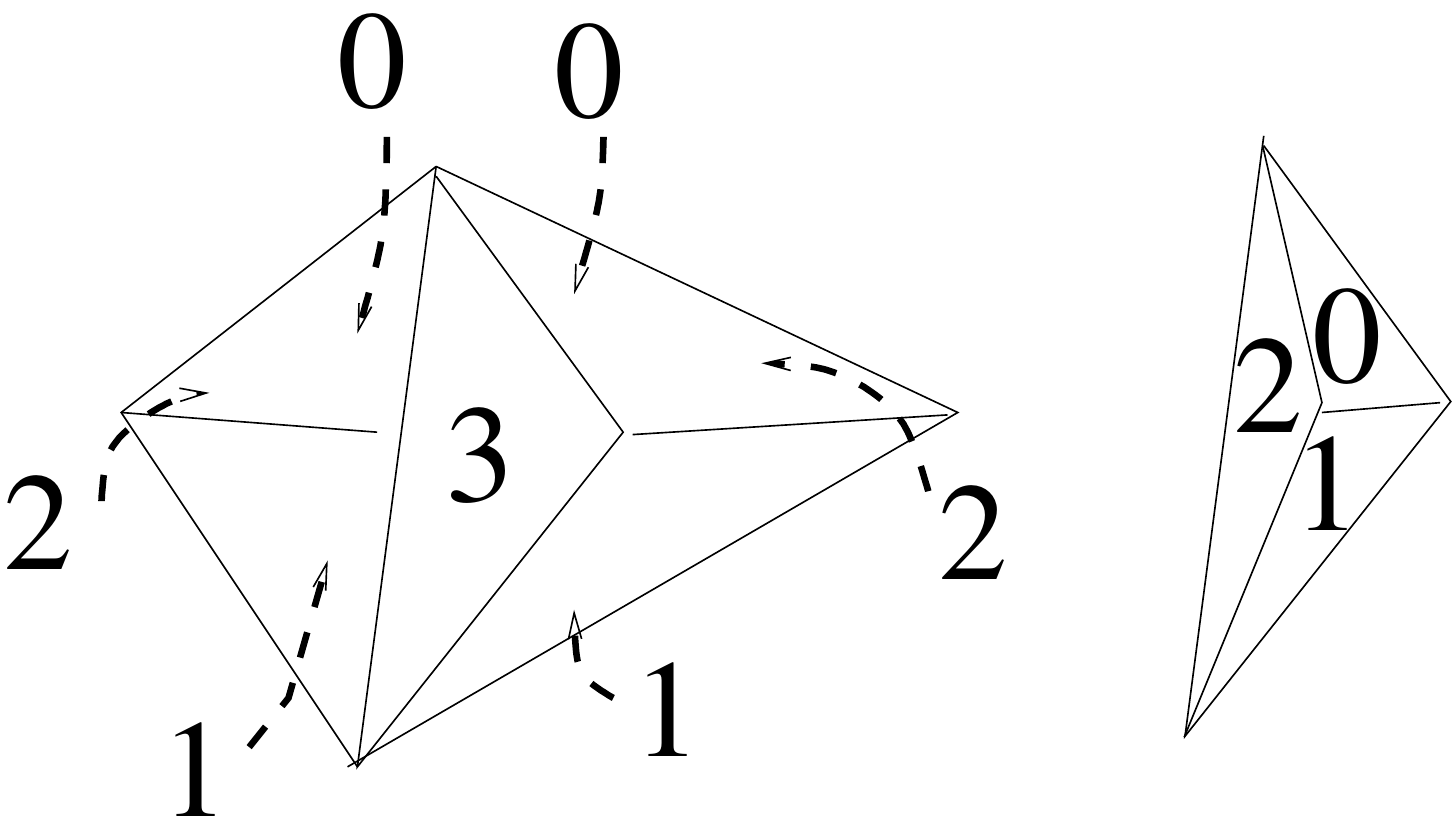}
\caption{1-Dipole contraction in $\cG$ and its dual gluing.}
\label{fig:1canc}
\end{center}
\end{figure}

\noindent{\bf 1-Dipoles.} Consider a line of color $3$ with end vertices $v$ and $w$ (denoted $L^3_{vw}$) 
in a graph $\cG$. Call $a_0$ ($a_1$ and $a_2$) the end vertex of the line of 
color $0$ ($1$ and $2$) touching $v$, and
$b_0$ ($b_1$ and $b_2$) the end vertex of the line of color $0$ ($1$ and $2$) 
touching $w$ (see figure \ref{fig:1canc}).
The vertices $v$ and $w$ belong each to some 3-bubble of colors $012$,
$B^{012}_{va_0a_1a_2} $ and $B^{012}_{wb_0b_1b_2}$. The two bubbles
might coincide or might be different. If they are different and at least one of them is
planar then the line $L^3_{vw}$ is called an {\it 1-Dipole}.

A 1-Dipole can be contracted, that is the line $L^3_{vw}$ together with 
the vertices $v$ and $w$ can be deleted from the graph and the remaining lines reconnected 
{\it respecting the coloring} (see figure  \ref{fig:1canc}).  
In the dual gluing a 1-Dipole of color $3$ represents two tetrahedra sharing the triangle 
(of color 3) such that the vertices opposite to the triangle (duals to $B_{va_0a_1a_2}^{012}$ and 
$B_{wb_0b_1b_2}^{012}$) are different. The contraction translates 
in squashing the two tetrahedra, merging the two vertices,
and coherently identifying the remaining triangles $0$, $1$ and $2$ 
(see figure \ref{fig:1canc}).

In this picture it is clear why one of $B_{va_0a_1a_2}^{012}$ or 
$B_{wb_0b_1b_2}^{012}$ is required to be planar. If both points opposite to the 
triangle $3$ were isolate singularities, the squashing of tetrahedra would decrease the 
number of singular points and would not be a homeomorphism. It is however a homeomorphism 
as long as one of the points is regular\footnote{See \cite{FG}, especially the remark 
on page 93 in the proof of the main theorem.}.

The vertices $v$ and $w$ belong to {\it the same} faces $03$, $13$ and $23$  
($F^{03}_{vwa_0b_0}$, $F^{13}_{vwa_1b_1}$, $F^{23}_{vwa_2b_2}$), but
{\it distinct} faces $01$, $02$ and $12$ ($F^{01}_{va_0a_1}$, $F^{01}_{wb_0b_1}$,
 $F^{02}_{va_0a_2 }$, $F^{02}_{wb_0b_2}$ and $F^{12}_{va_1a_2 }$, $F^{12}_{wb_2b_2}$).
They also belong to {\it the same} bubbles $013$, $023$ and $123$, 
($B^{013}_{vwa_0a_1b_0b_1}$, $B^{023}_{vwa_0a_2b_0b_2}$, $B^{123}_{vwa_1a_2b_1b_2}$)
but {\it different} bubbles $012$ ($B^{012}_{va_0a_1a_2}$ and $B^{012}_{wb_0b_1b_2}$).
We track the effect of the 1-Dipole contraction on the graph $\cG$.
Taking $B^{012}_{va_0a_1a_2}$ the planar bubble,  the contraction
\begin{itemize}
 \item deletes the vertices $v$ and $w$ and the line $L^3_{vw}$.
 \item glues $L^0_{va_0}$ on $L^0_{wb_0}$ to form a new line $L^0_{vw}$ (and similarly for colors $1$ and $2$).
 \item transforms the face $F^{03}_{vwa_0b_0}$ into a face $F^{03}_{a_0b_0}$ (and similarly for $13$ and $23$) .
 \item glues the face $F^{01}_{va_0a_1}$ on the face $F^{01}_{wb_0b_1}$ to form a new face $F^{01}_{a_0b_0b_1a_1}$
  (and similarly for $02$ and $12$).
 \item transforms the bubble $B^{013}_{vwa_0a_1b_0b_1}$ into a bubble $B^{013}_{a_0a_1b_0b_1}$ (and similarly for $023$ and $123$) 
 \item glues $B^{012}_{va_0a_1a_2}$ on $B^{012}_{wb_0b_1b_2}$ to form a new bubble $B^{012}_{a_0b_0a_1b_1a_2b_2}$.
\end{itemize}

The bubbles $013$, $023$ and $123$ transform trivially 
under contraction. Call $n$, $l$, $f$ and $g$ ($n'$, $l'$, $f'$ and $g'$)  
the vertices, lines, faces and genus of one of these bubbles before (after) 
contraction. We have
\bea
 |n'|=|n|-2 \;,\quad |l'|=|l|-3 \;, \quad |f'| = |f| -1 \Rightarrow g'=g \; .
\eea

The bubble $B_{va_0a_1a_2}^{012}$ (with $n_a$, $l_a$ $f_a$ and $g_a$) is 
glued on $B_{wb_0b_1b_2}^{012} $ (with $n_b$, $l_b$ $f_b$ and $g_b$) to form 
the new bubble $B^{012}_{a_0b_0a_1b_1a_2b_2}$ (with $n_b'$, $l_b'$, $f_b'$ and 
$g_b'$) and
\bea
&& |n_b'|= |n_a|+|n_b|-2 \; , \quad |l'_b| = |l_a|+|l_b| - 3 \; , \quad |f'_b|= |f_a|+ |f_b| -3
\Rightarrow g'_b= g_a+g_b \; .
\eea
Thus $g'_b=g_b$ if $g_a=0$. If $B_{wb_0b_1b_2}^{012} $ is dual to 
a conical singularity ($g_b\neq 0 $) then the new bubble $B^{012}_{a_0b_0a_1b_1a_2b_2}$
is dual to an identical singularity and the two dual pseudo manifolds are 
homeomorphic \cite{FG}. Were we to allow a contraction when both $g_a,g_b \neq 0$ we would
merge two conical singularities into a unique (more degenerate) conical singularity.

\bigskip

\noindent{\bf Amplitude.} Suppose that all lines enter $v$ and exit $w$. 
We denote $h_{0;v}$ the group element associated to $L^0_{va_0}$, etc. and 
use the shorthand notation $(01);v$ for $F^{01}_{va_0a_1}$ etc. 
The contribution of all faces containing $v$ and/or $w$
to the amplitude of $\cG$ is
\bea \label{eq:ampliii}
&& \int dh_{0;v} dh_{0;w} dh_{1;v} dh_{1;w} dh_{2;v} dh_{2;w} dh_3 \crcr
&& \delta^N_{(03)}( h_{0;v} h_3^{-1} h_{0;w} f^{03}  ) \;
\delta^N_{(13)}( h_{1;v} h_3^{-1} h_{1;w} f^{13}  ) \;
\delta^N_{(23)}( h_{2;v} h_3^{-1} h_{2;w} f^{23}  ) \crcr
&&\delta^N_{(01);v} (h_{0;v}h_{1;v}^{-1} f^{01}_v ) \;
\delta^N_{(02);v} ( h_{2;v} h_{0;v}^{-1} f^{02}_v ) \;
\delta^N_{(12);v} (h_{1;v}h_{2;v}^{-1} f^{12}_v) \crcr
&&\delta^N_{(01);w} (h_{1;w}^{-1} h_{0;w} f^{01}_w ) \;
\delta^N_{(02);w} ( h_{0;w}^{-1} h_{2;w}  f^{02}_w) \;
\delta^N_{(12);w} (h_{2;w}^{-1} h_{1;w} f^{12}_w) \; ,
\eea
where $f^{03}$ denotes the product of the remaining group elements along the face $03$ and 
similarly for the rest.
We first change variables to $ h_{0;w}'= h_3^{-1} h_{0;w} $, 
$d h_{0;w}' = d h_{0;w} $ (and similarly for  $h_{1;w}$ and $h_{2;w}$). 
The integral over $h_3$ is trivial. Forgetting the primes we obtain 
\bea
&& \int dh_{0;v} dh_{0;w} dh_{1;v} dh_{1;w} dh_{2;v} dh_{2;w} \crcr
&& \delta^N_{(03)}( h_{0;v}  h_{0;w} f^{03}  )  \;
\delta^N_{(13)}( h_{1;v}  h_{1;w} f^{13}  ) \; 
\delta^N_{(23)}( h_{2;v}  h_{2;w} f^{23}  ) \crcr
&&\delta^N_{(01);v}(h_{0;v}h_{1;v}^{-1} f^{01}_v ) \; 
\delta^N_{(02);v}( h_{2;v} h_{0;v}^{-1} f^{02}_v ) \;
\delta^N_{(12);v}(h_{1;v}h_{2;v}^{-1} f^{12}_v) \crcr
&&\delta^N_{(01);w} (h_{1;w}^{-1} h_{0;w} f^{01}_w ) \;
\delta^N_{(02);w} ( h_{0;w}^{-1} h_{2;w}  f^{(02)}_w) \;
\delta^N_{(12);w} (h_{2;w}^{-1} h_{1;w} f^{(12)}_w) \; .
\eea
We change again variables to $h_0 =  h_{0;v}  h_{0;w}$ $dh_0 = h_{0;w}$ (and similarly for $h_{1;w}$ and 
$h_{2;w}$) to obtain
\bea
 && \int dh_{0;v} dh_{0} dh_{1;v} dh_{1} dh_{2;v} dh_{2} \crcr
&& \delta^N_{(03)}( h_{0} f^{03}  ) \;
\delta^N_{(13)}( h_{1} f^{13}  ) \;
\delta^N_{(23)}( h_{2} f^{23}  ) \crcr
&&\delta^N_{(01);v} (h_{0;v}h_{1;v}^{-1} f^{01}_v ) \;
\delta^N_{(02);v}( h_{2;v} h_{0;v}^{-1} f^{02}_v ) \;
\delta^N_{(12);v} (h_{1;v} h_{2;v}^{-1} f^{12}_v) \crcr
&& \delta^N_{(01);w} (h_1^{-1}h_{1;v} h_{0;v}^{-1} h_0 f^{01}_w ) \;
 \delta^N_{(02);w} ( h_0^{-1}h_{0;v} h_{2;v}^{-1} h_2   f^{02}_w) \;
 \delta^N_{(12);w} ( h_2^{-1}h_{2;v} h_{1;v}^{-1} h_1 f^{12}_w) \; .
\eea
We integrate $h_{1;v}, h_{2;v} $ using $\delta^N_{(01);v}$ and $\delta^N_{(02);v}$ (hence  
$ h_{1;v} = f^{01}_v h_{0;v} \; , h_{2;v}^{-1} = h_{0;v}^{-1} f^{02}_v $ ) and eq. 
(\ref{eq:ampliii}) becomes 
\bea\label{eq:1dip}
 && \int dh_{0;v} dh_{0} dh_{1} dh_{2} \crcr
&& \delta^N_{(03)}( h_{0} f^{03}  ) \;
\delta^N_{(13)}( h_{1} f^{13}  ) \;
\delta^N_{(23)}( h_{2} f^{23}  ) \crcr
&& \delta^N_{(12);v} (f^{01}_v f^{02}_v f^{12}_v) \\
&& \delta^N_{(01);w} (h_1^{-1} f^{01}_v h_0 f^{01}_w ) \;
 \delta^N_{(02);w} ( h_0^{-1} f^{02}_v h_2   f^{02}_w) \; \delta^N_{(12);w} ( h_2^{-1} f^{12}_v h_1 f^{12}_w) \; .
\nonumber
\eea
Remark that, ignoring $\delta^N_{(12);v}$, the integrand of eq. (\ref{eq:1dip})
corresponds to the graph with the 1-Dipole contracted. But $\delta^N_{(12);v}$ 
reproduces the external face of a ribbon graph obtained by cutting the vertex $v$ in the bubble
$B^{012}_{va_0a_1a_2}$. The latter is a planar ribbon graph hence by eq. (\ref{eq:routingf}) 
$\delta^N_{(12);v} (f^{01}_v f^{02}_v f^{12}_v)$ can be replaced by $\delta^N(e)$. 
Recalling that the number of vertices decreases by $2$ we obtain that the amplitudes 
of $\cG$ and $\cG-L^3_{vw}$ (the graph with the 1-Dipole $L^3_{vw}$ contracted), 
are proportional 
\bea
 A^{\cG} = \frac{(\lambda \bar \lambda)} {\delta^N(e)} \delta^N(e)  A^{\cG-L^3_{vw}} =
 (\lambda \bar \lambda) A^{\cG-L^3_{vw}}
\; .
\eea

\bigskip

\begin{figure}[htb]
\begin{center}
 \includegraphics[width=5cm]{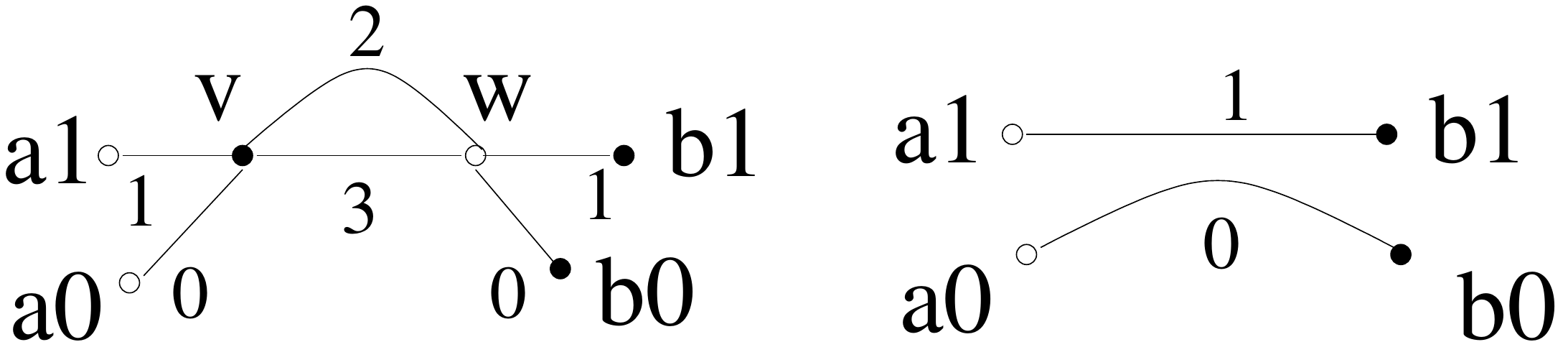}
\caption{2-Dipole contraction.}
\label{fig:2canc}
\end{center}
\end{figure}

\noindent{\bf 2-Dipoles.} A {\it 2-Dipole} of colors $23$ (see figure \ref{fig:2canc}) is a couple of lines connecting 
the same two vertices $v$ and $w$, $L^2_{vw}$ and $L^{3}_{vw}$ such that the 
faces $F^{01}_{va_0a_1}$ and $F^{01}_{wb_0b_1}$ are different. The 2-Dipole forms a face 
$F^{23}_{vw}$. Like the 1-Dipoles, the 2-Dipoles can be contracted (by deleting the lines $2$ and $3$ forming the 
2-Dipole and reconnecting the rest of the lines respecting the colors). 
This is represented in figure \ref{fig:2canc}.
After contraction the two faces $F^{01}_{va_0a_1}$ and $F^{01}_{wb_0b_1}$ are glued into a unique
face $F^{01}_{a_0a_1b_0b_1} $. A 2-Dipole is dual to two tetrahedra sharing two 
triangles (of colors 2 and 3 for figure \ref{fig:2canc}) such that the edge opposite to the two triangles 
in each tetrahedron (dual to the faces $F^{01}_{va_0a_1}$ and $F^{01}_{wb_0b_1}$) are different. 
The contraction translates in squashing the two tetrahedra and coherently identifying the 
remaining boundary triangles. This move always represents a homeomorphism \cite{FG}. 
Denoting $\cG-F^{23}_{vw}$ the graph obtained from $\cG$ after contracting the 2-Dipole, 
a short computation along the lines of the
one for 1-Dipoles yields
\bea
 A^{\cG}= \frac{(\lambda \bar\lambda)}{\delta^N(e)} A^{\cG- F^{23}_{vw}} \; .
\eea

The Dipole contraction moves can be inverted into Dipole creation moves. 
The fundamental result we will use in the sequel \cite{FG} is that two pseudo manifolds
dual to colored graphs $\cG$ and $\cG'$ are homeomorphic if $\cG$
and $\cG'$ are related by a finite sequence of 1 and 2-Dipole creation and 
contraction moves. We call two such graphs ``equivalent'', $\cG \sim \cG'$.

\section{Bubble routing and Core Graphs}\label{sec:multigems}

In the literature one finds two classes of results (bounds and evaluations) 
for amplitudes of GFT graphs. They are expressed either in terms of the number of 
vertices (\cite{sefu1,sefu2}) or in terms of the number of bubbles \cite{FreiGurOriti,sefu3}. 
In order to build the $1/N$ expansion in CGFT we need to strike the right balance
between the vertices and the bubbles of a graph. This is achieved by a bubble routing algorithm.

\bigskip

\noindent{\bf Bubble routing.} We start by choosing a set of roots of $\cG$ for all colors $i$. 
For the color $3$, if all the bubbles $\cB^{012}$ are planar we chose one of them as root and denote it $R^{012}_1$. 
If there exist non planar bubbles $012$ we set a non planar bubble as principal root $R^{012}_1$, and the other 
non planar bubbles as ``branch roots'' $R^{012}_2,R^{012}_3, \dots$. We denote the set of $012$ roots of 
$\cG$ by $\cR^{012}=\{ R^{012}_1,R^{012}_2,\dots\}$. We repeat this for all colors 
(and denote $\cR_{\cG}$ the set of all roots of $\cG$).

\begin{figure}[htb]
\begin{center}
 \includegraphics[width=4cm]{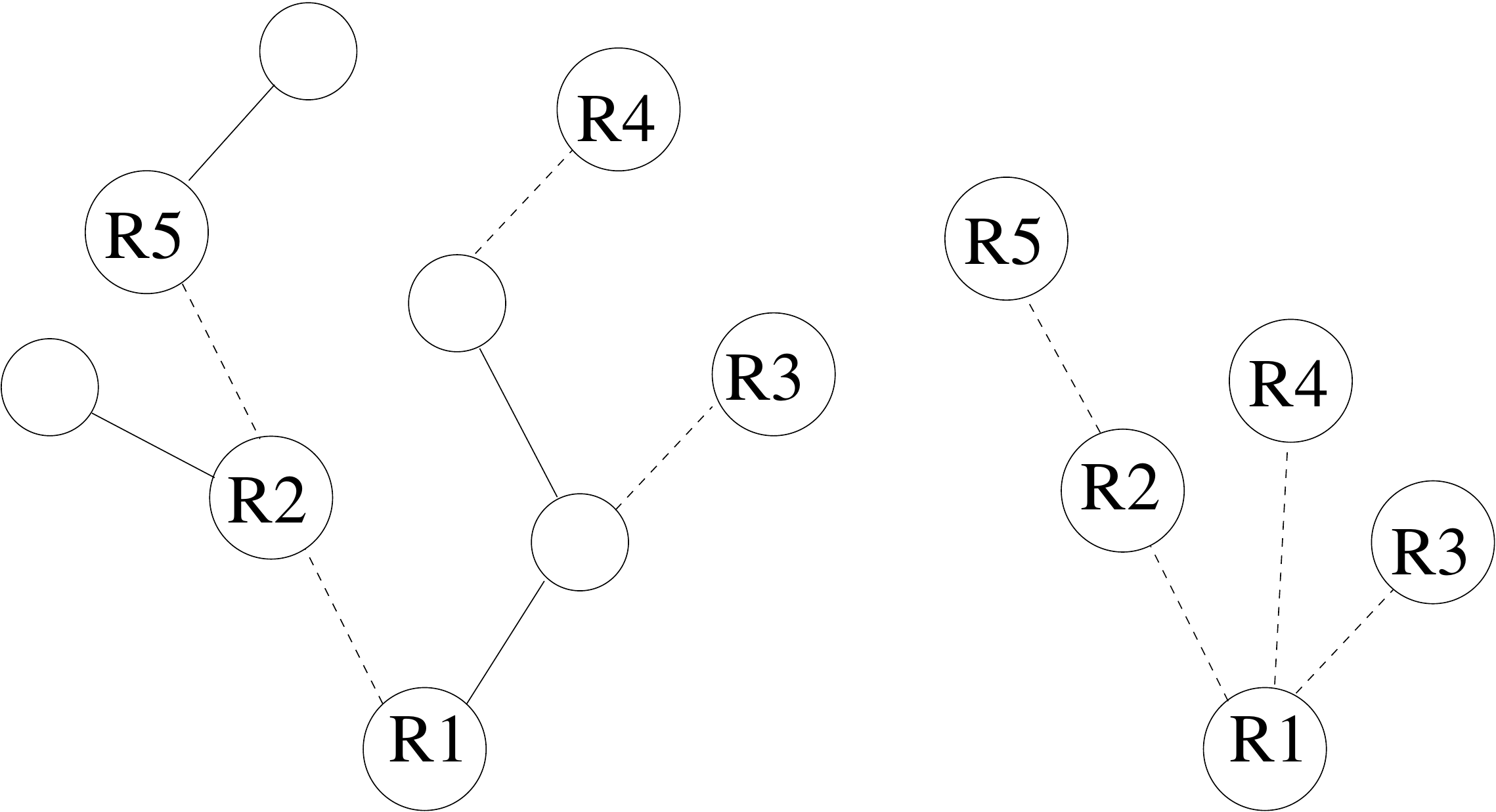}
\caption{A tree $\cT^3$ in the $012$ connectivity graph.}
\label{fig:treebub}
\end{center}
\end{figure}

We associate to the bubbles $012$ of $\cG$ a ``$012$ connectivity graph''. Its vertices represent
the various bubbles $012$. Its lines are the lines of color $3$ in $\cG$. 
They either start and end on the same bubble $012$ (in which case they are ``tadpole'' lines 
in the connectivity graph), or not. A particularly simple way to picture the $012$ connectivity
graph is to draw $\cG$ with lines $0$, $1$ and $2$ much shorter than the lines $3$.
We chose a tree in the connectivity graph, $\cT^{3}$ (and call the rest of the lines $3$ 
``loop lines''). For a branch root $R^{012}_q$, the line incident on it and belonging to the path in 
$\cT^{3}$ connecting $R^{012}_q$ to the principal root $R^{012}_1$ is represented as dashed.
All the other lines in $\cT^{3}$ are represented as solid lines. An example is given in 
figure \ref{fig:treebub}.

All the solid lines in $\cT^3$ are 1-Dipoles and we contract them. We end up with 
a connectivity graph with vertices corresponding to the roots $R_q^{012}$. The remaining 
lines of color $3$ cannot be 1-Dipoles (they are either tadpole lines or they 
separate two non planar roots). The number of 1-Dipoles of color $3$ contracted is 
$|\cB^{012}|-|\cR^{012}|$. Neither the number nor the topology of the bubbles of
the other colors, $B^{013}$, $B^{023}$ and $B^{123}$ is changed under these contractions.

Having exhausted a complete set of 1-Dipoles of color $3$, we repeat the procedure for the 1-Dipoles of color $2$.
The routing tree $\cT^2$ is chosen in the graph obtained {\it after} contracting the 1-Dipoles of color $3$ and
depends on $\cT^3$, $\cT^2(\cT^3)$. The contraction of 1-Dipole of color $2$ changes the 012 connectivity 
graph but it {\it cannot} create new 1-Dipoles of color $3$: the topology of the $012$ bubbles is unaffected by 
reducing 1-Dipoles of color $2$, hence the lines of color $3$ will still either be tadpole lines or separate two non 
planar roots $012$. After a full set of 1-Dipole contractions indexed by four distinct routing trees
$\cT^3, \cT^2(\cT^3),\cT^1(\cT^2,\cT^3),\cT^0(\cT^1,\cT^2,\cT^3)$ we obtain a 
{\it Core Graph}\footnote{If $\cG$ is dual to a manifold and one further reduces a full set of 2-Dipoles
one recovers a ``gem'' graph of \cite{Lins}.}.

\begin{definition}[{\bf Core Graph}]
   A colored graph with $2p$ vertices $\cG_p$ is called a Core Graph at order $p$ if, for all colors $i$,
   it either has a unique (planar or non planar) bubble $P^{\widehat{i} }_1$ or all its bubbles 
   $ P^{\widehat{i} }_1, P^{\widehat{i} }_2, \dots$ are non planar. 
\end{definition}

The amplitude of the graph $\cG$ and of the Core Graph obtained after routing are related by
\bea\label{eq:fam}
 A^{\cG}= (\lambda \bar\lambda)^{ |\cB_{\cG}|- |\cR_{\cG}| } A^{ \cG_{p} } \;, 
  \qquad 2p=| \cN_{\cG} | - 2 ( |\cB_{\cG}|- |\cR_{\cG}| ) \; .
\eea

The Core Graph one obtains by routing is {\it not} independent of the routing trees
$\cT^3, \cT^2,\cT^1,\cT^0$. The same graph leads to several equivalent Core Graphs, all at the same
order $p$, $\cG_p \sim \cG_p'\sim \dots$. One can prove that all equivalent Core Graphs 
at the same order $\cG_p \sim \cG_p'$ have the same amplitude.
Only the creation/contraction of dipoles of color $i$ can change the number of bubbles of 
colors $\widehat{i}$, and the latter only create/annihilate planar bubbles. It follows that 
the numbers of bubbles of colors $\widehat{i}$ of $\cG_p$ and $\cG_{p'}$ are equal 
and consequently the total numbers of 1-Dipole creations and contractions are equal. 
As $\cG_p$ and $\cG_p'$ have the same number of vertices, the total numbers of 2-Dipole 
creations and contractions are also equal and $A^{\cG_p}=A^{\cG_p'}$.

We denote $ \mathfrak{G}_p=\{ [\cG_p] \} $ the set of equivalence classes of 
Core Graphs at order $p$ under the equivalence relation $\sim$. The amplitude
is a well defined function of the equivalence class $[\cG_p]$. 
Under an arbitrary routing any graph will fall in a unique equivalence class
$[\cG_p]$. The free energy of the colored Boulatov model admits a topological expansion in
Core Graphs classes
\bea\label{eq:free}
 F(\lambda, \bar\lambda) = \sum_{p=1}^{\infty} \sum_{[\cG_p]\in \mathfrak{G}_p } 
C^{[\cG_p]}(\lambda,\bar\lambda) A^{[\cG_p]} \; ,
\eea
where $C^{[\cG_p]}( \lambda,\bar\lambda )$ is a combinatorial factor counting all the graphs
routing to a Core Graph class at order $p$. The scaling with $N$ is entirely captured by the Core Graph amplitude $A^{[\cG_p]}$.
A Core Graphs class is dual to a specific pseudo manifold. Note however that the same pseudo manifold is represented by
an infinity of classes $[\cG_p]$ at higher and higher orders in $p$. 

Core Graphs are in three dimensions the appropriate generalization of the super rosettes 
of two dimensional GFT. The only ingredient missing at this point is some estimate of their amplitude.

\begin{theorem}[{\bf The Core Graph bound}]\label{thm:multig}
  The amplitude of a Core Graph at order $p$, $\cG_p$, with set of bubble ${\cal P}$ respects
\bea
|A^{ \cG_p}| \le 
(\lambda \bar\lambda)^{p}
[\delta^N(e)]^{- \frac{1}{3}p + \frac{1}{3} \sum_{b\in {\cal P} }(1- g_b)  +1 } \; .
\eea
\end{theorem}
\noindent{\bf Proof:} We denote the set of lines and faces of $\cG_p$ by $\cL$ and $\cF$. The amplitude of the Core Graph is 
\bea
A^{ \cG_{p} } = 
\frac{(\lambda\bar\lambda)^p}{[\delta^N(e)]^p} 
\int \prod_{\ell\in \cL } dh_{\ell} 
\prod_{f \in \cF } \delta^N_{f}(\prod_{\ell \in f} ^{\rightarrow} h_{\ell}^{\sigma^{\ell | f}} )\; .
\eea

Denote $\cJ^{ij}$ the jacket of $\cG_p$ with the faces faces $ij$ and $\widehat{ij}$ deleted. 
The idea is to use the jacket graph to integrate explicitly as many group elements as possible. 
Indeed, routing the faces of the jacket graph will associate a line to all (save one) of
its faces. When integrating all (save one) of the $\delta^N$ functions of the faces of the 
jacket graph will contribute $1$, as $\int dh \; \delta^N(h^{-1} \dots ) K(h)= K(\dots)$.
The effect of this integrations over the rest of the $\delta^N$ functions is exceedingly complicated to track. 
However we will not need to do it, as we will just use a naive bound $\delta^N(h)\le \delta^N(e)$ for all of them.
In detail
\bea
A^{ \cG_{p} } = 
\frac{(\lambda\bar\lambda)^p}{[\delta^N(e)]^p} 
\int \prod_{\ell\in \cL } dh_{\ell} \;
\Big{[} \prod_{f'\in \cF^{ij}\cup \cF^{\widehat{ij} } }
\delta^N_{f'}(\prod_{\ell\in f'}^{\rightarrow} h_{\ell}^{\sigma^{\ell | f'}} ) \Big{]}
\Big{[} 
\prod_{f \in \cJ^{ij} } \delta^N_{f}(\prod_{\ell\in f}^{\rightarrow} h_{\ell}^{\sigma^{\ell | f}} )
\Big{]}
\; ,
\eea
and routing the faces of the jacket graph via a tree $\tilde \cT$ in the dual graph of the jacket we get
\bea
A^{ \cG_{p} } = 
\frac{(\lambda\bar\lambda)^p}{[\delta^N(e)]^p} 
&&
\int \prod_{\ell\in \cL_{p} } dh_{\ell} \; 
\Big{[} \prod_{f'\in \cF^{ij}\cup \cF^{\widehat{ij} } }
\delta^N_{f}(\prod_{\ell \in f'}^{\rightarrow} h_{\ell}^{\sigma^{\ell | f'}} ) \Big{]} \crcr
&&\Big{[}
 \delta^N_{r} (\prod_{\ell \notin \tilde \cT }^{\rightarrow} h_{\ell}^{\sigma^{\ell | \cup_{f\in \cJ^{ij}} f} } )
 \prod_{f\in \cJ^{ij}, f\neq r} 
\delta^N_{f} \Big{(} h_{l(f,\tilde \cT)}^{\sigma^{l(f,\tilde \cT) | f} } 
(\prod_{\ell \neq l(f,\tilde \cT) }^{\rightarrow} h_{\ell}^{\sigma^{\ell | f}}) \Big{)}
\Big{]}
 \; . 
\eea
Each of the $\delta^N$ of the faces of the jacket can now be associated uniquely to a specific integral over 
some group element. For all the lines in $\tilde \cT$ we change variables to 
\bea
  \tilde h_{l(f,\tilde \cT)} =  h_{l(f,\tilde \cT)}^{\sigma^{l(f,\tilde \cT) | f} } 
(\prod_{\ell \neq l(f,\tilde \cT) }^{\rightarrow} h_{\ell}^{\sigma^{\ell | f}}) \; ,
\eea
and write (in sloppy notations)
\bea
A^{ \cG_{p} } = 
\frac{(\lambda\bar\lambda)^p}{[\delta^N(e)]^p} 
\int \prod_{\ell\in \cL_{p} \setminus  \tilde \cT } dh_{\ell}  \;
\prod_{ l\in \tilde \cT } d\tilde h_{l }
\Big{[}  \prod_{f'\in \cF_p^{ij}\cup \cF_p^{\widehat{ij} } } 
\delta^N_{f'}( \dots )  \Big{]} \delta^N_{r} ( \dots  ) 
\Big{[} \prod_{f\in \cJ^{ij}, f\neq r} \delta^N_{f} \Big{(} \tilde h_{l(f,\tilde \cT)} \Big{)} \Big{]} \; . 
\eea
Each $\delta^N$ in the last line integrates with its associated $ \tilde h_{l(f,\tilde \cT)}  $, and we get
\bea
 A^{ \cG_{p} } = 
\frac{(\lambda\bar\lambda)^p}{[\delta^N(e)]^p} 
\int \prod_{\ell\in \cL_{p} \setminus  \tilde \cT } dh_{\ell}  
\Big{[}\prod_{f'\in \cF^{ij}\cup \cF^{\widehat{ij} } } 
\delta^N_{f'}( \dots ) \Big{]} \delta^N_{r} ( \dots  )
\le \frac{(\lambda\bar\lambda)^p}{[\delta^N(e)]^p}  [\delta^{N}(e)]^{|\cF^{ij}|+|\cF^{\widehat{ij}}| +1 } \; .
\eea 
One can use any of the three jackets of the graph to derive a bound. Using the jacket which yields the 
best estimate we always have
\bea 
A^{\cG_p} \le \frac{(\lambda\bar\lambda)^p}{[\delta^N(e)]^p}  [\delta^{N}(e)]^{\frac{|\cF|}{3} +1 } \; ,
\eea
and by eq. (\ref{eq:eule}) we have
\bea
 2p - 4 p + |\cF|- |{\cal P}| = - \sum_{b\in {\cal P}} g_b
\Rightarrow |\cF| =  2p+ \sum_{b\in {\cal P} } (1-g_b) \; .
\eea

\qed

Note that $\sum_{b\in {\cal P} } (1-g_b)\le 4 $ (and equal $4$ if and only if the Core Graph is dual to a manifold).
The Core Graph bound ensures that more and more complicated topologies (i.e. topologies which cannot be represented 
by a Core Graph with $p$ vertices or less) are suppressed at least as $[\delta^{N}(e)]^{\frac{7-p}{3}}$ in eq. 
(\ref{eq:free}).

\bigskip

\noindent{\bf The $1/N$ expansion.} We are now in the position to perform the $1/N$ expansion of the colored GFT model.
In order to evaluate all contributions to the order $[\delta^N(e)]^{-\alpha}$ one lists all (equivalence classes of)
Core Graphs up to order $p=3\alpha+7$. This is a finite problem, hence solvable. Then one computes the amplitude 
of each Core Graph (which can of course turn out to be much smaller than the value predicted by the Core Graph 
bound). The free energy is
\bea
 F(\lambda, \bar\lambda) = \sum_{p=1}^{3\alpha+7} 
\sum_{[\cG_p]\in \mathfrak{G}_p } C^{[\cG_p]}(\lambda,\bar\lambda) A^{[\cG_p]}  +O( [\delta^N(e)]^{- \alpha} ) \; .
\eea

The Core Graphs graphs up to $p=3$ are the graphs $\cG_1$, $\cG_2$, $\cG_{3,a}$, $\cG_{3,b}$, $\cG_{3,c}$ and 
$\cG_{3,d}$ from figure \ref{fig:graphs}. The Core Graphs $\cG_1$, $\cG_2$, $\cG_{3,a}$ and $\cG_{3,b}$ are dual to 
the three sphere $S^{3}$. The Core Graphs $\cG_{3,a}$ and $\cG_{3,b}$ are in the same equivalence 
class at order $3$. The Core Graphs $\cG_{3,c}$ and $\cG_{3,d}$ are dual to pseudo manifolds: $\cG_{3,c}$ has two non planar 
bubbles each of genus $1$, while $\cG_{3,d}$ has only one non planar bubble of genus $1$.
The Core Graph bound ensures that 
\bea
&&  A^{[\cG_1]} \le [\delta^{N}(e)]^{2}\;, \quad 
    A^{[\cG_2]} \le [\delta^{N}(e)]^{\frac{5}{3}} \;,\quad 
    A^{[\cG_{3,a}]}\le [\delta^{N}(e)]^{\frac{4}{3}}\; , \crcr 
&&
  A^{[\cG_{3,c}]} \le [\delta^N(e)]^{\frac{2}{3}} \;, \quad 
    A^{[\cG_{3,d}]} \le [\delta^N(e)] \; .
\eea 
Contributions coming from Core Graphs at higher order are at most of order $\delta^N(e)$. Direct computation shows that 
\bea
&&  A^{[\cG_1]} = [\delta^{N}(e)]^{2} \;,\quad  
    A^{[\cG_2]} = [\delta^{N}(e)] \;, \quad 
    A^{[\cG_{3,a}]} = [\delta^{N}(e)]^{0}  \; , \crcr
&&  A^{[\cG_{3,c}]} =  \frac{1}{\delta^N(e)} \int dhdu \delta^N(hu^{-1}h^{-1}u) \;, \quad 
    A^{[\cG_{3,d}]} = [\delta^N(e)]^0 \; .
 \eea

Hence the partition function of the colored Boulatov model develops as
\bea
  F(\lambda,\bar\lambda) = C^{[\cG_1]}(\lambda,\bar \lambda) [\delta^N(e)]^2 + O([\delta^N(e)]) \; ,
\eea
and all graphs contributing to the dominant order are dual to manifolds 
homeomorphic with the three sphere $S^3$.  

\section*{Acknowledgements}

Research at Perimeter Institute is supported by the Government of Canada through Industry 
Canada and by the Province of Ontario through the Ministry of Research and Innovation.


\begin{thebibliography}{99}

\bibitem{laurentgft}
  L.~Freidel,
  Int.\ J.\ Theor.\ Phys.\  {\bf 44}, 1769 (2005)
  [arXiv:hep-th/0505016].

\bibitem{quantugeom2}
  D.~Oriti,
  [arXiv:0912.2441 [hep-th]].

\bibitem{mm} 
 F.~David,
  Nucl.\ Phys.\  B {\bf 257}, 543 (1985).

\bibitem{mmgravity}
 M.~Gross,
  Nucl.\ Phys.\ Proc.\ Suppl.\  {\bf 25A}, 144 (1992).

\bibitem{ambj3dqg}
  J.~Ambjorn, B.~Durhuus and T.~Jonsson,
  Mod.\ Phys.\ Lett.\  A {\bf 6}, 1133 (1991).

\bibitem{sasa1}
  N.~Sasakura,
  Mod.\ Phys.\ Lett.\  A {\bf 6}, 2613 (1991).

\bibitem{GFT} 
 D.~V.~Boulatov,
  Mod.\ Phys.\ Lett.\  A {\bf 7}, 1629 (1992)
  [arXiv:hep-th/9202074].

\bibitem{FreidelLouapre}
  L.~Freidel and D.~Louapre,
  Class.\ Quant.\ Grav.\  {\bf 21}, 5685 (2004)
  [arXiv:hep-th/0401076].

\bibitem{gftnoncom}
  A.~Baratin and D.~Oriti,
  [arXiv:1002.4723 [hep-th]].

\bibitem{newmo2} 
  J.~Engle, R.~Pereira and C.~Rovelli,
  Nucl.\ Phys.\  B {\bf 798}, 251 (2008)
  [arXiv:0708.1236 [gr-qc]].

\bibitem{newmo3}
  E.~R.~Livine and S.~Speziale,
  Phys.\ Rev.\  D {\bf 76}, 084028 (2007)
  [arXiv:0705.0674 [gr-qc]].

\bibitem{newmo4}
   L.~Freidel and K.~Krasnov,
  Class.\ Quant.\ Grav.\  {\bf 25}, 125018 (2008)
  [arXiv:0708.1595 [gr-qc]].

\bibitem{newmo5}
  J.~B.~Geloun, R.~Gurau and V.~Rivasseau,
  arXiv:1008.0354 [hep-th].

\bibitem{sing}
 R.~Gurau,
  Class.\ Quant.\ Grav.\ {\bf 27}, 235023 (2010)
  arXiv:1006.0714 [hep-th].

\bibitem{Brezin:1977sv}
  E.~Brezin, C.~Itzykson, G.~Parisi and J.~B.~Zuber
  Commun.\ Math.\ Phys.\  {\bf 59}, 35 (1978).
  
\bibitem{Alexandrov:2010un}
  S.~Alexandrov and P.~Roche,
  arXiv:1009.4475 [gr-qc].

\bibitem{color}
 R.~Gurau,
  [arXiv:0907.2582 [hep-th]].

\bibitem{PolyColor}
  R.~Gurau,
  Annales Henri Poincare {\bf 11}, 565 (2010)
  [arXiv:0911.1945 [hep-th]].


\bibitem{FreiGurOriti} 
  L.~Freidel, R.~Gurau and D.~Oriti,
  Phys.\ Rev.\  D {\bf 80}, 044007 (2009)
  [arXiv:0905.3772 [hep-th]].

\bibitem{sefu1}
  J.~Magnen, K.~Noui, V.~Rivasseau and M.~Smerlak,
  Class.\ Quant.\ Grav.\  {\bf 26}, 185012 (2009)
  [arXiv:0906.5477 [hep-th]].

\bibitem{sefu2}
  J.~B.~Geloun, J.~Magnen and V.~Rivasseau,
  [arXiv:0911.1719 [hep-th]].

\bibitem{sefu3}
   J.~B.~Geloun, T.~Krajewski, J.~Magnen and V.~Rivasseau,
  Class.\ Quant.\ Grav.\  {\bf 27}, 155012 (2010)
  [arXiv:1002.3592 [hep-th]].

\bibitem{param}
  R.~Gurau and V.~Rivasseau,
  Commun.\ Math.\ Phys.\  {\bf 272}, 811 (2007)
  [arXiv:math-ph/0606030].

\bibitem{Lins} S.~Lins, {\it Gems, Computers and Attractors for 3-Manifolds, (Series on Knots and 
Everything, Vol 5) } ISBN: 9810219075/ ISBN-13: 9789810219079 

\bibitem{FG}
  M. ~Ferri and C.~Gagliardi
  Pacific\ Journal\ of\ Mathematics\ Vol. 100, No. 1, 1982

\bibitem{Bonzom:2010ar}
  V.~Bonzom and M.~Smerlak,
  Lett.\ Math.\ Phys.\  {\bf 93}, 295 (2010)
  [arXiv:1004.5196 [gr-qc]].

\bibitem{GW}
  H.~Grosse and R.~Wulkenhaar,
  Commun.\ Math.\ Phys.\  {\bf 256}, 305 (2005)
  [arXiv:hep-th/0401128].

\bibitem{GW1}
   R.~Gurau, J.~Magnen, V.~Rivasseau and F.~Vignes-Tourneret,
  Commun.\ Math.\ Phys.\  {\bf 267}, 515 (2006)
  [arXiv:hep-th/0512271].

\bibitem{GW2}
  V.~Rivasseau, F.~Vignes-Tourneret, R.~Wulkenhaar,
  Commun.\ Math.\ Phys.\  {\bf 262}, 565-594 (2006).
  [hep-th/0501036].

\end{thebibliography}
\end{document}